\documentclass[usenatbib]{mn2e}
\usepackage{natbib}
\usepackage{amssymb}
\usepackage{amsmath}

\newcommand{\HI}{H{\scriptsize{\textrm{I}}}}
\newcommand{\HII}{H{\scriptsize{\textrm{II}}}}
\newcommand{\HeI}{He{\scriptsize{\textrm{I}}}}
\newcommand{\HeII}{He{\scriptsize{\textrm{II}}}}
\newcommand{\HeIII}{He{\scriptsize{\textrm{III}}}}
\newcommand{\Lylim}{Lyman limit}
\newcommand{\fstar}{$f^*_{{\mathrm{esc}}}$}
\newcommand{\Lmin}{$L_{{\mathrm{min}}}$}
\newcommand{\zsix}{$z \sim 6$}
\newcommand{\zol}{$z_{\mathrm{ol}}$}
\newcommand{\omegab}{$\Omega_{\mathrm{B}}$}
\newcommand{\omegam}{$\Omega_{\mathrm{M}}$}
\newcommand{\omegal}{$\Omega_{\mathrm{\Lambda}}$}
\newcommand{\solarb}{$L_{\mathrm{{B}}}$}
\newcommand{\nul}{$\nu_{\scriptscriptstyle{\mathrm{L}}}$}
\newcommand{\nuz}{$\nu_{\scriptscriptstyle\mathrm{z}}$}
\newcommand{\mathnul}{\nu_{\scriptscriptstyle\mathrm{L}}}
\newcommand{\mathnuz}{\nu_{\scriptscriptstyle\mathrm{z}}}
\newcommand{\etanul}{$\eta_{\nu_{\scriptscriptstyle\mathrm{L}}}$}
\newcommand{\mathetanul}{\eta_{\nu_{\scriptscriptstyle\mathrm{L}}}}
\newcommand{\mathetanuz}{\eta_{\nu_{\scriptscriptstyle\mathrm{z}}}}

\newcommand{\mathLnul}{L_{\nu_{\scriptscriptstyle\mathrm{L}}}}
\newcommand{\deltacrit}{$\Delta_{\mathrm{{crit}}}$}

\def\simgt{\mathrel{\spose{\lower 3pt\hbox{$\sim$}}
        \raise 2.0pt\hbox{$>$}}}
\def\simlt{\mathrel{\spose{\lower 3pt\hbox{$\sim$}}\raise 2.0pt\hbox{$<$}}}
 
\title[The Quasar Contribution to the Reionisation]{Constraining the Quasar Contribution to the Reionisation of Cosmic Hydrogen}

\author[Srbinovsky \& Wyithe]{J.A. Srbinovsky \& J.S.B. Wyithe\\
School of Physics, University of Melbourne, Parkville, Victoria, Australia
 Email: jsrbino@physics.unimelb.edu.au}

\date{Draft Version}
\pagerange{\pageref{firstpage}--\pageref{lastpage}}
\pubyear{2006}

\def\LaTeX{L\kern-.36em\raise.3ex\hbox{a}\kern-.15em
    T\kern-.1667em\lower.7ex\hbox{E}\kern-.125emX}

\begin{document}

\label{firstpage}

\maketitle

\begin{abstract}
\noindent 
Absorption spectra of high redshift quasars suggest that the reionisation of cosmic hydrogen was complete near {\zsix}. The dominant sources of ionising photons responsible for this reionisation are generally thought to be stars and quasars. In this paper we make a quantitative estimate of the relative contributions made by these sources. Our approach is to compute the evolution of the post overlap ionising background radiation by combining semi-analytic descriptions of reionisation in a clumpy medium with a model for the quasar luminosity function. Our overall model has two free parameters, the star formation efficiency and the minimum quasar luminosity. 
By adjusting these parameters, we constrain the relative contributions made by stars and quasars through comparison with reported observations \citep{Fan2005}. We find that the relative quasar contribution (at $z=5.7$) to the ionising background was between $1.4\%$ and $14.5\%$. The range of uncertainty is dominated by the unknown minimum quasar luminosity.
\end{abstract}


\section{{INTRODUCTION}}
\label{introduction}

Observations of high redshift quasars imply that the reionisation of cosmic hydrogen was completed by {\zsix} [\citet[hereafter F05]{Fan2005}; \citet{Gnedin2006, White2003}]. It is commonly believed that reionisation was due to UV photons produced by the first stars and quasars \citep{Barkana}.
However, theoretical uncertainties and a lack of observational evidence make it difficult to estimate the relative contributions made by these sources.
Although quasars and stars contribute equally to the ionising background by $z\sim3$ \citep{Kriss2001,Smette2002}, it is often thought that stars dominated the UV flux at $z \gtrsim 6$ (e.g. Fan et al.~2001). 

Attempts to make quantitative estimates of the contribution of quasars to the end of the reionisation era have proceeded on several fronts and led to inconsistent results. 
For example, \citet[ hereafter MHR99]{Madau1999} establish a threshold photon emission rate at $z=5$ below which it would not be possible to maintain the reionisation of the inter-galactic medium (IGM). 
They then compute the photon emission rate coinciding with the known QSO population at $z\sim5$ and conclude that the quasar contribution to the budget of ionising photons was $\sim 10^{-2}$.
Analogously, \citet{Fanstripe2001} compute the quasar contribution to the photon emission rate at {\zsix} using the luminosity function (LF) computed from \textit{Sloan Digital Sky Survey} (\textit{SDSS}) observations of high redshift quasars.
They conclude that the quasar contribution to the ionising background at $z=6$ was also $\sim 10^{-2}$ and suggest that quasars could not be a major contributor of ionising photons at {\zsix} without invoking a large negative luminosity evolution in the LF.
Similar conclusions are drawn by \citet{Yan2004} and noted by \citet{Stiavelli2004} in support of the view that the major contribution to the ionising background at {\zsix} comes from star forming galaxies. A different approach was taken by \citet{Dijkstra2004}, who show that the unresolved component in the observed soft X-ray background is beneath the level that would be contributed by a population of quasars that could \emph{fully} ionise the universe at {\zsix}.

On the other hand \citet{Bunker2004} show that the contribution to the ionising background from star forming galaxies identified in the \emph{Hubble Ultra-Deep Field (\textit{UDF})} and \emph{Great Observatories Origins Deep Survey (\textit{GOODS})} is insufficient to produce reionisation at {\zsix}. The missing ionising flux could be generated by quasars, or it could be generated by dwarf galaxies which were suppressed following reionisation \citep{Wyithe2006}. Alternatively, \citet{Meiksin2005} finds that a quasar population described by a statistical fit to the double power law form of the LF \citep{Boyle1988} falls short of the required flux to reionise the universe at {\zsix} only by a factor of $2-3$ and could therefore contribute this missing ionizing flux.

This range of conclusions reflects the differing approaches used to describe the ionising sources and more significantly, the different photo-emission rates assumed to be necessary for reionisation. 
In the aforementioned analyses, reionisation in an inhomogeneous (\emph{clumpy}) universe was treated using a clumping factor, $C=\langle n ^2 \rangle /\langle n  \rangle^2 $ (where $n$ is the hydrogen number density), to account for the increased recombination rate. 
Larger values of $C$ increase the emission rate necessary to achieve reionisation. 
MHR99 consider a value of $C=30$ corresponding to the numerical simulation of \citet{Gnedin1997}, however they note the considerable uncertainty in the applicability of this value. \citet{Fanstripe2001}, \citet{Yan2004}, \citet{Stiavelli2004} and \citet{Bunker2004} use this same value for $C$. \citet{Dijkstra2004} use a value of $C=10$ and \citet{Meiksin2005} assumes values of $1<C<10$. 

Much of the uncertainty in the ionising photon budget therefore centers around the value of the clumping factor. It is important to note that the clumping factor should be computed over the volume of IGM that must be maintained in an ionised state. This volume should not include the collapsed regions inside the virial radius of dark matter halo hosts containing star forming galaxies (the additional flux required to keep this overdense region ionised is accounted for by the value of escape fraction of ionising photons). The volume over which the clumping factor should be computed also excludes those overdense regions in the IGM that form {\textit \Lylim} systems. 
These overdense clouds set the mean-free-path (MFP) for ionising photons and the recombination rate should be computed only up to densities separating the ionised IGM and these clouds. A framework within which the clumping factor appropriate for reionisation in an inhomogeneous IGM may be computed was described in \citet[ hereafter M-E00]{Miralda2000}. We follow the formalism developed in that work throughout this paper.

We begin by describing our models for the reionisation history and quasar LF in \S~\ref{models}. We then review our procedures for calculation of the ionising background using these models in \S~\ref{evaluating_flux} and  present results from our analysis in \S~\ref{results}, before discussing our conclusions in \S~\ref{conclusions}. Throughout the paper we assume values for cosmological parameters based on {\textit{WMAP3}} results \citep{Spergel2006} but obtained by averaging over the observational data sets from \textit{WMAP3, SDSS, HST, SN Astier \textrm{and} BAO}.
The resulting parameters are {\omegal}~$=.743$, {\omegam}~$=.257$, {\omegab}$=.0437$ and $h=.718$ (Renyue Cen, private communication). In computation of the mass function we assume a primordial power spectrum defined by a power law with index $n=.95$, an exact transfer function given by \citet{Bardeen1986} and rms mass density fluctuations with a sphere of radius $R_8=8h^{-1}\mathrm{Mpc}$ of $\sigma_8=0.765$.

\section{Semi-Analytic Models for Reionisation and the Quasar Luminosity Function}
\label{models}
The density field in an inhomogeneous universe can be described by the overdensity, $\Delta_i = \rho_i / \overline{\rho}$.
M-E00 show how the effective recombination rate in an inhomogeneous universe may be determined dynamically by consideration of the maximum overdensity ({\deltacrit}) penetrated by ionising photons within {\HII} regions (i.e. the volume of the IGM to be maintained in an ionised state).
\citet[ hereafter WL03]{Wyithe2003a} employed this prescription for the recombination rate into a semi-analytic model of reionisation. 
 In this model {\deltacrit} is fixed prior to overlap, after which it evolves with redshift [see equation (6) and relevant discussion in WL03]. 
In this formalism it is possible to compute the effective clumping factor in regions where $\Delta_i < {\mbox{\deltacrit} }$ such that $C=C[z,{\mbox{\deltacrit}(z)}]$. 
For example, assuming overlap to occur at $z\sim 6$ and {\deltacrit}$=20$ prior to overlap (following WL03), $C(z=6) \sim 2.3$. Thus, the effective clumping factor in ionised regions is considerably smaller than the value often assumed, which is over the whole IGM. 

The sources of ionising photons in this reionisation model were assumed to be quasars, in addition to  population II (popII) and population III stars. In the current work we limit our attention to reionisation due to popII stars (Gnedin \& Fan~2006), which govern the final stages of reionisation even in the presence of an earlier partial or full reionisation by population III stars (Wyithe \& Loeb~2003b).
The quasar LF model of Wyithe \& Loeb~(2003b) successfully reproduces\footnote{This model is less successful at reproducing the number of very luminous quasars at {\zsix} using the WMAP third year cosmology, indicating the necessity of revision of its physical basis. However we maintain the LF described in Wyithe \& Loeb~(2003b) as an empirically successful description.} observations of the quasar number density over a large range of luminosities at $z\gtrsim3$.

Our model includes two free parameters. First, the minimum luminosity for quasars {\Lmin}, which is not predicted by the model. Second, the parameter {\fstar} represents the product of star formation efficiency and escape fraction, and characterises the contributions made by stars. 
Our approach is to compute a reionisation history given a particular set of {\Lmin} and {\fstar}. With this history in place we then compute the evolution of the background radiation field due to these same sources.
Post overlap, ionising photons will experience attenuation due to residual overdense pockets of {\HI} gas. 
We use the reionisation model of WL03 to describe the ionising photon MFP, and subsequently derive the attenuation of ionising photons. We then compute the flux at the {{\textit \Lylim}}~({\nul}~$=3.29\times10^{15}\mathrm{Hz})$ in the IGM due to sources immediate to each epoch, in addition to redshifted contributions from earlier epochs. 

We note that {\HI} {\Lylim} photons $(13.6\hspace{2pt}eV)$ are incapable of ionising helium (He) [$24.6\hspace{2pt}eV$ ({\HeII}), $54.4\hspace{2pt}eV$ ({\HeIII})]. We therefore neglect He when computing the intensity of the ionising background. 
However, the presence of He is fully incorporated in the reionisation model of WL03 which we use to describe the history of reionisation.

\section{{EVALUATING FLUX AT THE LYMAN LIMIT}}
\label{evaluating_flux}

In this section we review the process for calculating the ionising background flux at a particular redshift. We assume the background to be generated by a combination of stars and quasars. We first define $z_0$ to be the redshift at which the flux is to be evaluated. The flux at the Lyman limit is normalised in \emph{physical} units of  $J_{21}$  $(10^{-21} \mathrm{ergs/sec/Hz/cm^2/sr})$, and at redshift $z_0$ is related to the energy density by

\begin{equation}
\label{dEdV_to_flux}
J_{21}(z_0) = \frac{c}{4\pi}\hspace{2pt}\frac{d^2E_{\mathnul}^{{\mathrm{tot}}}(z_0)}{dV{d\nu}}\frac{1}{10^{-21}}
(1+z_0)^3,
\end{equation}

\noindent
where $c$ is the speed of light and $\frac{ d^2 E_{{\mbox{\scriptsize\nul}}}^{{\mathrm{tot}}}(z_0) } {dV{d\nu}}$ is the energy per unit frequency interval, per co-moving volume, at frequency {\nul}, and $z_0$.
Contributions to the radiation field at {\nul} (and $z_0$) from sources at redshift $z$ were emitted at frequency {\nuz}~ $= \frac{1+z}{1+z_0}${\nul}.
Note that {\nuz} remains below the He {\Lylim} $(24.6\hspace{2pt} eV)$ whilst $\frac{1+z}{1+z_0} \la 2$.
The semi-analytic model of the reionisation history predicts an average redshift of overlap, which we refer to as {\zol}. While the observed redshift of overlap is subject to some scatter among different lines-of-sight, we make the approximation that the universe becomes transparent to Ly$\alpha$ photons at {\zol} along all lines-of-sight. 

The ionising background flux contains contributions from sources at $z_0$ in addition to redshifted flux from sources at higher redshift.
To compute the ionising background flux we consider contributions from sources with redshifts $z_0<z<z_{\mathrm{ol}}$, and therefore assume the complete attenuation of ionising photons emitted prior to the redshift of overlap. 
The total co-moving energy density at redshift $z_0$ is
{
\begin{equation}
\label{fluxeq}
\frac{d^2E_{{\mbox{\scriptsize\nul}}}^{\mathrm{tot}}(z_0)}{dVd\nu}=\int_{z_{\mathrm{ol}}}^{z_0}dz 
\frac{d^3E_{{\mbox{\scriptsize\nuz}}}(z)}{dVd{\nu}dt}
e^{-\tau(z,z_0)}
\frac{dt}{dz},
\end{equation}
}

\noindent
where the exponential term describes the attenuation of ionising photons (\S~\ref{attenuation}) between redshifts $z$ and $z_0$.
In this equation one might expect to find factors of $\left(\frac{1+z_0}{1+z}\right)$ due to expansion. However, energy loss due to redshifting of the photon frequency is cancelled by the redshifting of the frequency interval.
The term $\frac{d^3E_{{\mbox{\scriptsize\nuz}}}(z)}{dVd{\nu}dt}$ represents the frequency dependent (co-moving) energy density per unit time generated by ionising sources. Finally, the relation between proper time and redshift is
\begin{equation}
\label{redshift-time}
\frac{dt}{dz} 
=
[H_0(1+z)
\sqrt{(1+z)^3 {\mbox {\omegam}}+{\mbox {\omegal}}}]
^{-1} ,
\end{equation}
where $H_0$ is the current value of the Hubble parameter. Equations~(\ref{dEdV_to_flux})~\&~(\ref{fluxeq}) describe the ionising flux at $z_0$ given contributions of ionising sources at higher redshifts. In the next two sub-sections we describe the calculation of the emissivity of quasars and stars, which provide the sources of ionising radiation.

\subsection{{{\Lylim} photons from Quasars}}
\label{quasar_flux}
Our estimate of the emissivity of quasars is based on the B-band quasar LF [$\Phi({\mbox{\solarb}},z)$], i.e. the number density per co-moving volume per solar B-band luminosity {\solarb}. 
We use the theoretically derived LF of \citet{WyitheSMBH2003}, which successfully reproduces known properties of the observed quasar LF at both the bright and faint ends at $z \gtrsim 2.5$, including the isolated observations discussed in {\S~\ref{constraining}}. 
This model is based on the physics of galaxy mergers, accretion powered luminosity and the self-regulated growth of super-massive black holes. The total B-band luminosity $[{\mbox {\solarb}}^{\mathrm{tot}}(z)]$ per volume is calculated directly from $\Phi({\mbox{\solarb}},z)$
{
\begin{equation}
\label{Ltot_integral}
\frac{d{\mbox{\solarb}}^{\mathrm{tot}}(z)} {dV}
= \int _{\mbox{\Lmin}}^{\infty} d{\mbox\solarb} \hspace{1pc} \Phi ({\mbox{\solarb}},z)\times{\mbox \solarb}.
\end{equation}
}
\noindent While this integral converges at high luminosities, the shallow slope of the LF at low luminosities means that the value of the integral is sensitive to the lower limit {\Lmin}, which is highly uncertain.
We therefore treat {\Lmin} as a free parameter in our description of quasar emissivity.
To convert from {\solarb} to the luminosity at the {\Lylim} $({\mathLnul})$ we use the relation described by \citet{Schirber2003},

\begin{equation}
{\mathetanul}=\frac{\mathLnul}{\mbox\solarb}=10^{18.05} \mathrm{ergs\hspace{2pt}s^{-1}  Hz^{-1}  {\mbox\solarb}^{-1}}.
\end{equation}
In order to compute the {\Lylim} flux at $z_0$, we require the contribution from intrinsically higher frequency photons from higher redshifts. We assume a powerlaw spectrum {$L_\nu\propto\nu^{-\alpha}$}, whence we obtain
\begin{equation}
\label{Schirber_relation}
\mathetanuz=\mathetanul\frac{\phi(\mathnuz)}{\phi(\mathnul)}=\mathetanul\left( \frac{1+z}{1+z_0}\right)^{-\alpha}.
\end{equation}
\citet{Telfer} find $\alpha=1.57$ in the wavelength range $500\lesssim \lambda \lesssim 1200$~\AA~ using (radio quiet) AGN spectra obtained from the \textit{HST} UV survey.
We adopt this value which is also used by \citet{Schirber2003} in determining {\etanul}. 
In this work we compute the contribution of type-I quasars to reionisation. There may be additional accreting objects whose ionizing flux is blocked by a dusty torus \citep{Elitzur2006}. In unified schemes these obscured objects are the same as type-I quasars, however they are viewed from a different orientation. Our quasar LF accounts for the observed luminosity density of ionising radiation. We assume that all ionising photons escape the host galaxy [note however that the Ly$\alpha$ halos around some QSOs indicate reprocessing of this ionising flux \citep{Francis2006}].
Finally, using equation~(\ref{fluxeq}), we find the energy density due to quasars at frequency {\nul} and redshift $z_0$,  

\begin{align}
\label{quasar_energy_density}
\frac{d^2E_{\mathnul}^{\mathrm{tot}}(z_0)}{dVd\nu} =&  \hspace{2pt}{\mathetanul}\int^{z_{0}}_{z_{\mathrm{ol}}}dz \hspace{1pt}  
\frac{ d{\mbox\solarb}^{\mathrm{tot}}(z)}{dV}
\hspace{2pt}\left(\frac{1+z_{0}}{1+z}\right)^{{\alpha}} \notag\\
& \times e^{-\tau(z,z_0)}\frac{dt}{dz}.
\end{align}
\noindent
This energy density may be converted to a flux and expressed in units of $J_{21}$ using equation \ref{dEdV_to_flux}.

\subsection{{\Lylim} photons from Stars}
\label{stellar_flux}
We next describe the contribution made by popII stars. To begin, we describe the spectral energy distribution (SED) of popII star forming galaxies using the model presented in \citet{Leitherer}. The SED has units of { 
$\frac{d^3E_\nu}{d\nu dtd\dot{M}} $}, where {$\dot{M}$} is expressed in units of (baryonic) solar masses per year. In the instance that ionising photons are produced primarily in starbursts, with lifetimes much shorter than the Hubble time, we may express the star formation rate per unit time as
\begin{equation}
\frac{d\dot{M}}{dV}(z)=f^*\frac{dF(z)}{dt_{\mathrm{year}}}\hspace{2pt}\rho_{\mathrm{b}},
\end{equation}
where $\rho_{\mathrm{b}}$ is the co-moving baryonic mass density, and $F(z)$ is the collapsed fraction of mass in halos above a critical mass at $z$. The factor $f^*$ (star formation efficiency) describes the fraction of collapsed matter that participates in star formation. This fraction is largely unknown, however \citet{Wyithe2006} constrain $f^*$ to be $\sim 10-15\%$ by fitting to the galaxy LF at high redshift.

To compute $F(z)$ we use the Press-Schechter~(1976) mass function with the modification of \citet{Sheth} to describe the evolution of the collapsed fraction [F(z)] within halos above a critical virial temperature. In a cold neutral IGM beyond the redshift of reionisation, this critical virial temperature is set by 
the temperature ($T_{{\mathrm{N}}}\sim 10^4$ K) above which efficient atomic hydrogen cooling promotes star formation. Following the reionisation of a region, the Jeans mass in the heated IGM limits accretion to halos above $T_{{\mathrm{I}}}\sim 2.5\times10^5$ K \citep{thoul}. The corresponding virialised mass can be determined using equation (26) in \citet{Barkana}. 

We may therefore write the time derivative of the collapsed fraction 
\begin{align}
\label{collapsed_fraction}
\frac{dF}{ dt_{  {\mathrm{year}}  }}(z)= & 
\left[ Q_{\mathrm{m}}(z)\frac{dF(z,T_{{\mathrm{I}}})}{dz} + [1-Q_{\mathrm{m}}(z)]
\frac{dF(z,T_{\mathrm{N}})}{dz}\right] \notag\\
& \times \frac{dz}{dt_{{\mathrm{year}}}},
\end{align}

\noindent
here $Q_{\mathrm{m}}$ is the ionised baryonic mass fraction in the universe. 

To describe the ionising flux from stars we require one further parameter. 
We have assumed that all ionising photons produced by quasars escape the host galaxy (\S~\ref{quasar_flux}). However, due to the softer spectra of galaxies it is likely that only a fraction of ionising photons produced by stars enter the IGM. 
Therefore, an additional factor of $f_{\mathrm{esc}}$ (the escape fraction) must be included when computing the emissivity due to stars.
\citet{Ciardi2005} include a review of existing constraints on $f_{\mathrm{esc}}$ which suggests that its value is $\la 15\%$. The star formation efficiency and escape fraction may be combined into a single free parameter ({\fstar}) to describe the contribution of stars in our model.

Finally the energy density due to popII stars at $z_0$ may be computed using equation~\ref{fluxeq} with $\frac{d^3E_{\nu_z}(z)}{dVd{\nu}dt}$  given by
{
\label{stellar_energy_density}
\begin{equation}
\frac{d^3E_{\mbox\nuz}(z)}{dVd{\nu}dt}\hspace{2pt} 
=
\frac{d^3E_{\mbox\nuz}(z)}{d\nu dt{d\dot{M}}}
\frac{dF(z)}{dt_{\mathrm{year}}}\hspace{2pt}\rho_{\mathrm{b}} {\mbox\fstar}.
\end{equation}
}
\noindent This energy density may then be converted to a flux and combined with the contribution due to quasars.

\subsection{\normalsize{Attenuation of ionising photons}}
\label{attenuation}

Given the luminosity density we may find the ionising background using equation (2), following calculation of the optical depth ($\tau$) 
and thereby the attenuation of ionising photons.
Following overlap, we use the approach of M-E00 to estimate the MFP ($\lambda_i$) of ionising photons as a function of redshift $(z_i)$.

\begin{equation}
\label{Miralda-EscudeMFP}
{\lambda}_i ={\lambda}_0(1-F_{\mathrm{v}})^{-{2}/{3}}.
\end{equation}
Here we use the reionisation model of WL03 to compute the redshift evolution of the the volume filling factor in ionised regions. 
\begin{equation}
\label{volume_fraction}
F_{\mathrm{v}} = \int_0^{{\mbox\deltacrit}(z)}d{\Delta} P({\Delta}) ,
\end{equation}
where [$P(\Delta)$] is the volume weighted probability distribution of the overdensity (M-E00).
The product ${\lambda}_0 H=60 \hspace{1pt}\mathrm{kms}^{-1}$ was obtained from comparison to the scales of $Ly\alpha$ forest structures in simulations at $z=3$ (M-E00). 
In reality, the MFP is set by residual {\HI} in ionisation equilibrium with the background flux as well as {\Lylim} systems. 
Our estimate of the MFP therefore provides an upper limit.
Furthermore, our reionisation model assumes ionising photons to be instantly absorbed, an assumption which becomes less applicable post overlap. 

Finally the MFP as a function of redshift may be used to compute the attenuation of ionising photons between redshifts $z$ and $z_0$.
The resulting optical depth is

\begin{equation}
\label{optical_depth}
\tau(z,z_0) = \int^{z_0}_{z}dz
\frac{cdt}{dz}\frac{1}{\lambda_i}.
\end{equation}

\section{RESULTS }
\label{results}
\begin{figure*}
\vspace*{120mm}
\includegraphics{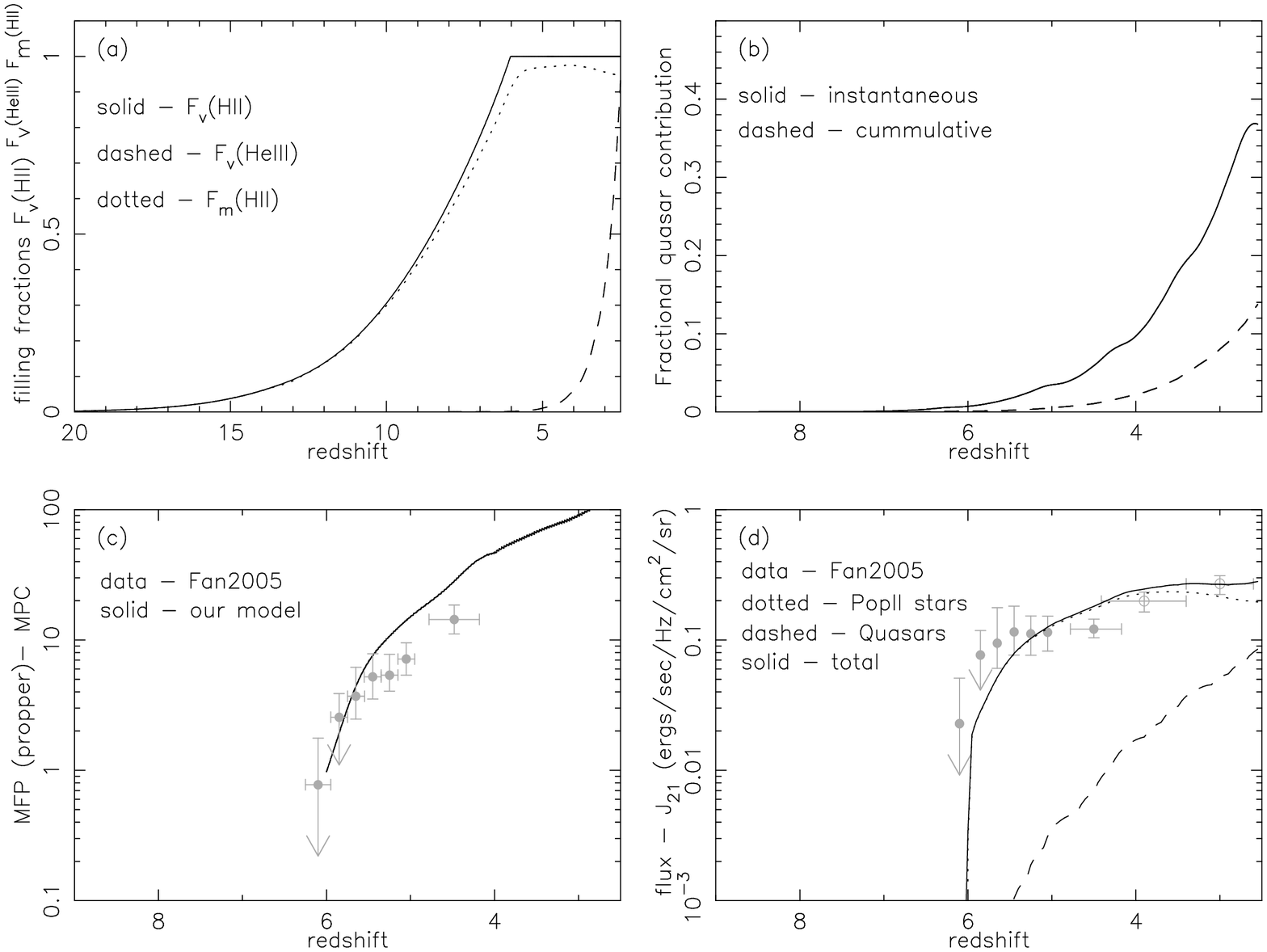}
\caption{\label{fig1} The {\textit{best fit}} model (see text) corresponding to {\fstar}$=.00815$, {\Lmin}$=10^{12.5}${\solarb} and {\deltacrit}=20. 
(a) The reionisation history. Solid, dotted and dashed lines represent the evolution of the volume and mass filling fractions [$F_{\mathrm{v}}$({\HII}),$F_{\mathrm{m}}$({\HII}),$F_{\mathrm{v}}$({\HeIII})].
(b) The fractional contribution of quasars to the instantaneous production rate of ionising photons (solid line). 
The fractional contribution of quasars to the cumulative number of ionising photons produced by {\textit{z}} (dashed line). 
(c) The evolution of the MFP. Our model (solid line) and observational data points from  F05.
(d) The redshift evolution of $J_{21}$. The model combined flux (solid line) and contributions from stellar sources (dotted line) and quasars (dashed line). 
The data points are from F05. 
}
\label{4panels}
\end{figure*}

The model described in the previous section has two free parameters {\Lmin} and {\fstar}. For any combination of these parameters we are able to compute a reionisation history, the evolution of the MFP, the evolution of the ionising background, and the relative contributions of stars and quasars to the ionising background. In this section we first summarise the observations to which we compare our model before describing the results of this comparison.

\subsection{Observational estimates of the ionising background and mean-free-path}

F05 have analyzed the absorption spectra of 19 high redshift quasars. Based on this analysis F05 present estimates of the neutral fraction at a range of redshifts near the end of the reionisation era, as well as the background ionising flux and MFP. In this paper it is our aim to constrain the contributions of quasars and stars to the ionising background based on these observations. 

The observed estimates of the ionising background are quoted as photoionisation rates (F05), derived from the observed $(Ly{\alpha})$ optical depth. 
These photoionisation rates may be related to ionising background fluxes using

\begin{equation}
\label{Gamma}
\Gamma=
4\pi
\int_{\mbox\nul}^{\infty}
\frac{J_{\nu}}{h_{\mathrm{p}}\nu}
\sigma_{\nu}
d\nu ,
\end{equation}

\noindent where $\sigma_{\nu}$ is the {\HI} photoionisation \textit{cross-section} 
and $h_{\mathrm{p}}$ is Planck's constant.
Using equation~(\ref{Gamma}) and the spectra describing our sources, we find $\Gamma_{12}=1.69J_{21}$ (where $\Gamma_{12}$ is the ionising rate in units of $10^{-12}\mathrm{s}^{-1}$) for a background in which stars are the dominant source. Similarly, if the background radiation is dominated by quasars we obtain $\Gamma_{12}=2.64J_{21}$. In comparing our model to observation we normalise the observed flux using a weighted sum of the instantaneous flux contributions from each of the stellar and quasar sources.

F05 also estimate the ionising photon MFP. They compute MFPs using the frequency averaged cross-section ($\langle\sigma_{\nu}\rangle$) to {\Lylim} absorption assuming an ionising background spectrum of the form $J_{\nu}\propto\nu^{-\alpha}$.
They compute $\langle\sigma_{\nu}\rangle$ in the frequency range between the {\HI} and the {\HeI} {\Lylim}s,
and assume $\alpha=5$ which describes an ionising background dominated by stars \citep{Barkana}. 
In \S~\ref{stellar_flux}, we outlined our calculation of stellar emissivity using the spectrum of \citet{Leitherer}.
We therefore use this spectrum to adjust the values of $\langle\sigma_{\nu}\rangle$ from F05
(assuming stellar dominance) and present data points using this modified value.

\subsection{Comparison with observation}
\label{comparison}

In this section we show results for a particular model in which we assume ${\mbox\Lmin}=10^{12.5}${\solarb}.
In this case we find that the model best fits the ionising background observations for the value of {\fstar}$=.00815$. 
For these parameters we find that model fit to observations of the ionising background yields a reduced $\chi^2$ value of 1.3.
In a subsequent section, we explore the range of values for {\Lmin} permitted by observation.

For our representative model, Figure 1 shows the redshift evolution of four quantities computed.
Here we have assumed a critical overdensity of {\deltacrit $=20$} prior to overlap, which occurs in this case at ${\mbox\zol} \sim 6$.
Figure 1(a) shows the full reionisation history.
The volume fraction of {\HII} (solid line), the mass fraction of {\HII} (dotted line) and the volume filling fraction of doubly ionised {\HeIII} (dashed line) are shown. Note that overlap is marked by the point when the volume filling fraction of {\HII} reaches unity.
Figure~1(b) shows the resulting fractional contribution of quasars to the instantaneous production rate (solid line) of ionising photons. 
The dashed line shows the fractional contribution of quasars to the cumulative number of ionising photons produced by redshift $z$.
Ionising photons above the {\HeI} {\Lylim} also contribute to the reionisation of {\HI} (WL03). We therefore consider the contribution of ionising photons made by quasars and stars at all frequencies above the {\HI} Lyman limit.
The contribution of quasars to the instantaneous production rate rises towards low redshift and becomes $\sim 40\%$ that of stars by $z \sim 2.5$, in broad agreement with observation \citep{Kriss2001,Smette2002}. Cumulatively the stellar contribution remains dominant even at low redshifts.

\begin{figure*}
\vspace*{80mm}
\includegraphics{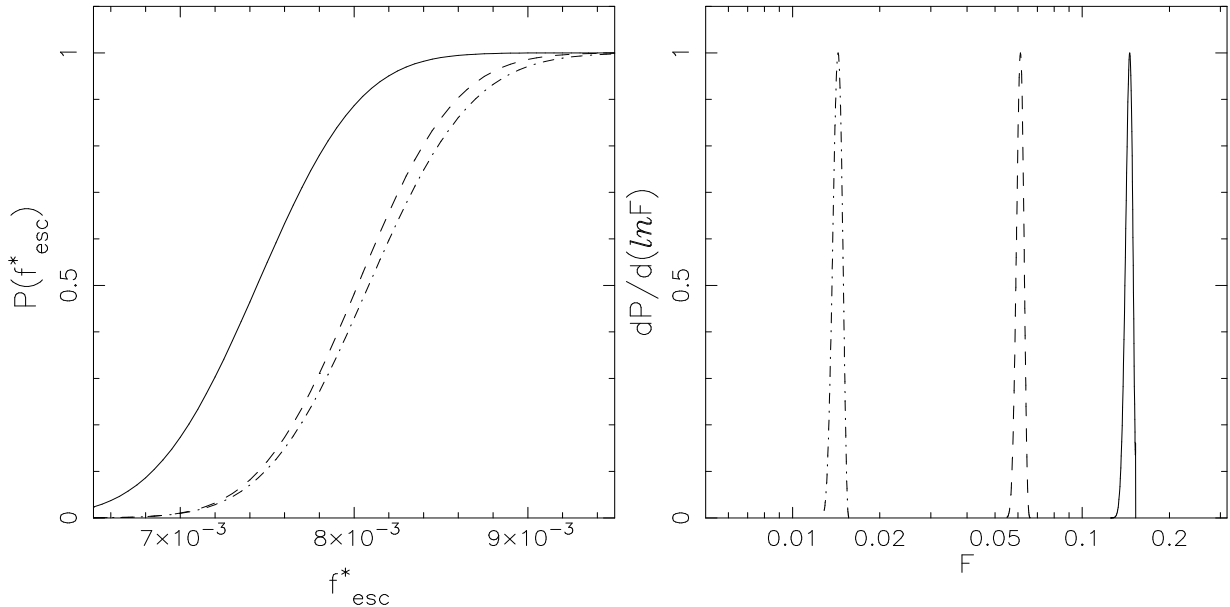}
\caption{\label{fig3}{The {\textit{a posteriori}} probability (P) assuming values for {\Lmin}({\solarb}) $=10^{11}$~(solid), $10^{11.75}$~(dashed), $10^{12.5}$~(dot-dashed). 
(a) The cumulative distribution as a function of {\fstar}. 
(b) $dP/dF$ where $F$ is the fractional quasar contribution to the instantaneous photon production rate at $z = 5.7$.}}
\label{4Lmins}
\end{figure*}

In Figure 1(c) we show the evolution of the MFP (solid line) with comparison to the points from F05. It is clear from this plot that 
the models abilty to predict the MFP diminishes towards lower redshift.
In Figure 1(d), we plot $J_{21}$ as a function of redshift. The evolution of the total (model) flux due to quasars and stars is shown as a solid line. The contribution to the ionising background due to quasars (stars) alone is shown as a dashed line (dotted line).
In Figures (c) and (d), excepting the two highest redshift bins the data points show the mean value for $J_{21}$ with $1-\sigma$ error bars.
In the two highest redshift bins optical depth measurements were not always possible due to complete \emph{Gunn-Peterson} troughs along some lines of sight (see F05).
In these cases the mean and standard deviation were calculated using lower limits on the optical depth. 
Hence these data points represent upper limits on $J_{21}$.

\subsection{{{\textbf{Constraining}} {\fstar} and the quasar flux contribution to the IGM }}
\label{constraining}
In the previous sub-section we discussed the results from a particular model which assumed a value for {\Lmin}$=10^{12.5}${\solarb}.
In this sub-section we investigate values of {\fstar}, given various minimum quasar luminosities.
In particular, for a range of {\Lmin} we compute the \textit{a posteriori} probability of our model flux, as a function of the free parameter {\fstar}, given the {\textit{likelihoods}} of our model fits. We include an \textit{a priori} probability distribution which is flat in the logarithm and consider a range for {\fstar} $\in[.006,.01]$. 

At low redshift $(z\la .5)$ quasars are observed at luminosities corresponding to $\sim10^{10.7}{\mbox\solarb}$ \citep{Boyle2000}.
\citet{Barger2003} have inferred from X-ray observations a quasar with luminosity of $\sim10^{11.3}{\mbox\solarb}$  at ${z=5.7}$. 
At $z=5.85$ observations have revealed the presence of quasars at luminosities corresponding to $\sim10^{12.8}{\mbox\solarb}$ \citep{Cool2006}.
This suggests an upper limit to {\Lmin} as even at high redshift we can see these quasars directly. 
At $z\sim4.5$ \citet{Dijkstra2006} find a downturn in the density of AGN with $L<10^{11}{\mbox\solarb}$, based on the absence of AGN in Ly$\alpha$ surveys.
However, current data is insufficient to strongly constrain the lower limit of QSO activity. Therefore, within this range we construct conditional probability distributions at {\Lmin}({\solarb}) ${= 10^{11}, 10^{11.75}, 10^{12.5}}$. 
In each case probability distributions are extracted from the aforementioned likelihood analysis.
For display we normalise the probability density functions such that the maximum likelihood is set to one.

In  Figure~{\ref{4Lmins}}(a) we show the resulting cumulative \textit{a posteriori} conditional probability distribution for {\fstar}. The models which best fit the ionising background observations, for each given {\Lmin}, have corresponding values for {\fstar} in the narrow interval $.0074 \lesssim \mbox\fstar \lesssim .008$. Furthermore, for each particular {\Lmin} the distribution is restricted to a narrow interval in {\fstar} space. This is a reflection of the sensivity of the goodness of fit to {\fstar}.
The best fits for each {\Lmin} correspond to overlap redshifts suggested by the reionisation model at $z \sim 6$.
This is  slightly lower than the recent numerical analysis of Gnedin \& Fan~(2006) which suggested overlap occurs between $6.1 \lesssim \mbox\zol \lesssim 6.3$. Better fits are obtained for larger values of $L_{\rm min}$. However, this reflects systematic uncertainty in the model rather than a real constraint. In Figure~{\ref{4Lmins}}(b) we show the corresponding distribution $\frac{dP}{dF}= \frac{dP} {d{f^*_{\mathrm{esc}}}}\frac{d{f^*_{\mathrm{esc}}}}{dF}$, 
where $F$ represents the fractional contribution of quasars to the instantaneous photon production rate at $z = 5.7$. 
For the values of {\Lmin} assumed in Figure~{\ref{4Lmins}} the most probable (fractional) quasar contributions to the ionising background are  ${\mathrm{F}\sim .145, .061, .014}$. This range covers the systematic uncertainty introduced by {\Lmin}. 

The models presented thus far have assumed a value of ${\mbox\deltacrit}=20$ beyond {\zol}.  We have repeated our analysis for different choices of {\deltacrit} and find that while the results are similar, the fit to $J_{21}$ improves as {\deltacrit} is increased between 5 and 20.
In particular, for higher {\deltacrit} the flux immediately following overlap rises more sharply due to both the increased {\fstar} that is required to match observation and also the higher value of $F_{\mathrm{m}}$ at {\zol}.

\subsection{Additional Uncertainties}
\label{additional_uncertainties}
Our model contains several additional uncertainties. These include, but are not limited to, the following. First, the model LF describes the density of high redshift quasars. The uncertainty in the overall level of quasar flux at $z\sim6$ (where there is only 1 point) is at least a factor of 2. This uncertainty needs to be added to the systematic uncertainty in {\Lmin}. Second, the MFP computed by our model is an upper limit. Indeed comparison of our model with the results of Fan et al.~(2005) suggest an overestimate by a factor of $\sim2$ at $z \lesssim 5.5$. This overestimate results in an underestimate of the attenuation in our model. We may therefore underestimate the quantity {\fstar} required for a good description of the ionising background, and therefore overestimate the fractional contribution to quasars. Thirdly, our model does not consider the possible evolution of {\Lmin} with redshift.

\section{CONCLUSIONS}
\label{conclusions}
In this paper we have estimated the relative contributions made by quasars and stars to the ionising background at the end of the reionisation epoch.
We have characterised these contributions by two free parameters. {\Lmin} which defines the minimum luminosity of contributing quasars and {\fstar} which describes the product of star formation efficiency and escape fraction of ionising photons from galaxies. We compute the relative quasar contribution for the range of values, {\Lmin}({\solarb})~$=10^{11},10^{11.75},10^{12.5}$. We find the relative contributions made by quasars to the ionising background at $z=5.7$ are ${\mathrm{F}\sim .145, .061, .014}$, respectively. The value of {\Lmin} provides the greatest uncertainty.

We are able to quantitatively constrain the contribution of quasars to the ionising background because our model (following M-E00) computes the clumping factor in the IGM which has represented a major uncertainty in previous studies.
There are a few independent checks which enhance our confidence in the results. 
Firstly, our quasar LF accurately reproduces observations at $z>2.5$. This quasar population was responsible for the double reionisation of helium, which for values of $L_{\rm min}(L_{\rm B})\sim10^{11}$ is predicted by our reionisation model to occur at $3\lesssim z \lesssim 3.5$, in agreement with observation  \citep{Bernardi2003,Theuns}. Secondly, the contribution of quasars to the ionising background becomes comparable to that of stars by $z \sim 2.5$ \citep{Kriss2001,Smette2002}. 
Thirdly, our model predicts {\fstar}$=.00815$ which is consistent with the product of the values $f^* \sim 10-15\%$~\citep{Wyithe2006} and $f_{esc} \lesssim 15\%$.~\citep{Ciardi2005} which are estimated from other studies. 
In future, improvements to the analysis presented here could be made by adding a population of quasars to cosmological simulations of reionisation (e.g. Gnedin \& Fan~2006) which more accurately reproduce the evolution of the ionising background.

\section*{Acknowledgements}
This work was supported in part by the Australian Research Council. JAS acknowledges the support of an Australian Postgraduate Award.

\bibliographystyle{mn2e}
\bibliography{flux_paper}

\begin{thebibliography}{}

\bibitem[\protect\citeauthoryear{{Bardeen}, {Bond}, {Kaiser} \&
  {Szalay}}{{Bardeen} et~al.}{1986}]{Bardeen1986}
{Bardeen} J.~M.,  {Bond} J.~R.,  {Kaiser} N.,    {Szalay} A.~S.,  1986, \apj,
  304, 15

\bibitem[\protect\citeauthoryear{{Barger}, {Cowie}, {Capak}, {Alexander},
  {Bauer}, {Brandt}, {Garmire} \& {Hornschemeier}}{{Barger}
  et~al.}{2003}]{Barger2003}
{Barger} A.~J.,  {Cowie} L.~L.,  {Capak} P.,  {Alexander} D.~M.,  {Bauer}
  F.~E.,  {Brandt} W.~N.,  {Garmire} G.~P.,    {Hornschemeier} A.~E.,  2003,
  \apjl, 584, L61

\bibitem[\protect\citeauthoryear{{Barkana} \& {Loeb}}{{Barkana} \&
  {Loeb}}{2001}]{Barkana}
{Barkana} R.,  {Loeb} A.,  2001, \apjs, 349, 125

\bibitem[\protect\citeauthoryear{{Bernardi}, {Sheth}, {SubbaRao}, {Richards},
  {Burles}, {Connolly}, {Frieman}, {Nichol}, {Schaye}, {Schneider}, {Vanden
  Berk}, {York}, {Brinkmann} \& {Lamb}}{{Bernardi} et~al.}{2003}]{Bernardi2003}
{Bernardi} M.,  {Sheth} R.~K.,  {SubbaRao} M.,  {Richards} G.~T.,  {Burles} S.,
   {Connolly} A.~J.,  {Frieman} J.,  {Nichol} R.,  {Schaye} J.,  {Schneider}
  D.~P.,  {Vanden Berk} D.~E.,  {York} D.~G.,  {Brinkmann} J.,    {Lamb} D.~Q.,
   2003, \aj, 125, 32

\bibitem[\protect\citeauthoryear{{Boyle}, {Shanks}, {Croom}, {Smith}, {Miller},
  {Loaring} \& {Heymans}}{{Boyle} et~al.}{2000}]{Boyle2000}
{Boyle} B.~J.,  {Shanks} T.,  {Croom} S.~M.,  {Smith} R.~J.,  {Miller} L.,
  {Loaring} N.,    {Heymans} C.,  2000, \mnras, 317, 1014

\bibitem[\protect\citeauthoryear{{Boyle}, {Shanks} \& {Peterson}}{{Boyle}
  et~al.}{1988}]{Boyle1988}
{Boyle} B.~J.,  {Shanks} T.,    {Peterson} B.~A.,  1988, \mnras, 235, 935

\bibitem[\protect\citeauthoryear{{Bunker}, {Stanway}, {Ellis} \&
  {McMahon}}{{Bunker} et~al.}{2004}]{Bunker2004}
{Bunker} A.~J.,  {Stanway} E.~R.,  {Ellis} R.~S.,    {McMahon} R.~G.,  2004,
  \mnras, 355, 374

\bibitem[\protect\citeauthoryear{{Ciardi} \& {Ferrara}}{{Ciardi} \&
  {Ferrara}}{2005}]{Ciardi2005}
{Ciardi} B.,  {Ferrara} A.,  2005, Space Science Reviews, 116, 625

\bibitem[\protect\citeauthoryear{{Cool}, {Kochanek}, {Eisenstein}, {Stern},
  {Brand}, {Brown}, {Dey}, {Eisenhardt}, {Fan}, {Gonzalez}, {Jannuzi},
  {McKenzie}, {Rieke}, {Soifer}, {Spinrad} \& {Elston}}{{Cool}
  et~al.}{2006}]{Cool2006}
{Cool} R.~J.,  {Kochanek} C.~S.,  {Eisenstein} D.~J.,  {Stern} D.,  {Brand} K.,
   {Brown} M.~J.~I.,  {Dey} A.,  {Eisenhardt} P.,  {Fan} X.,  {Gonzalez} R.~F.,
   {Jannuzi} B.~T.,  {McKenzie} E.~H.,  {Rieke} M.,  {Soifer} B.~T.,  {Spinrad}
  H.,    {Elston} R.~J.,  2006, astro-ph/0605030

\bibitem[\protect\citeauthoryear{{Dijkstra}, {Haiman} \& {Loeb}}{{Dijkstra}
  et~al.}{2004}]{Dijkstra2004}
{Dijkstra} M.,  {Haiman} Z.,    {Loeb} A.,  2004, \apj, 613, 646

\bibitem[\protect\citeauthoryear{{Dijkstra} \& {Wyithe}}{{Dijkstra} \&
  {Wyithe}}{2006}]{Dijkstra2006}
{Dijkstra} M.,  {Wyithe} J.~S.~B.,  2006, astro-ph/0606334

\bibitem[\protect\citeauthoryear{{Elitzur} \& {Schlosman}}{{Elitzur} \&
  {Schlosman}}{2006}]{Elitzur2006}
{Elitzur} M.,  {Schlosman} I.,  2006, astro-ph/0605686

\bibitem[\protect\citeauthoryear{{Fan}, {Narayanan}, {Lupton} \&
  {Strauss}}{{Fan} et~al.}{2001}]{Fanstripe2001}
{Fan} X.,  {Narayanan} V.~K.,  {Lupton} R.~H.,    {Strauss} M.~A.,  2001, \aj,
  122, 2833

\bibitem[\protect\citeauthoryear{{Fan}, {Strauss}, {Becker}, {White}, {Gunn},
  {Knapp}, {Richards}, {Schneider}, {Brinkmann} \& {Fukugita}}{{Fan}
  et~al.}{2005}]{Fan2005}
{Fan} X.,  {Strauss} M.~A.,  {Becker} R.~H.,  {White} R.~L.,  {Gunn} J.~E.,
  {Knapp} G.~R.,  {Richards} G.~T.,  {Schneider} D.~P.,  {Brinkmann} J.,
  {Fukugita} M.,  2005, astro-ph/0512082

\bibitem[\protect\citeauthoryear{{Francis} \& {McDonnell}}{{Francis} \&
  {McDonnell}}{2006}]{Francis2006}
{Francis} P.~J.,  {McDonnell} S.,  2006, Monthly Notices of the Royal
  Astronomical Society, 370, 1372

\bibitem[\protect\citeauthoryear{{Gnedin} \& {Fan}}{{Gnedin} \&
  {Fan}}{2006}]{Gnedin2006}
{Gnedin} N.~Y.,  {Fan} X.,  2006, astro-ph/0603794

\bibitem[\protect\citeauthoryear{{Gnedin} \& {Ostriker}}{{Gnedin} \&
  {Ostriker}}{1997}]{Gnedin1997}
{Gnedin} N.~Y.,  {Ostriker} J.~P.,  1997, \apj, 486, 581

\bibitem[\protect\citeauthoryear{Kriss, Shull, Oegerle, Zheng, Davidsen,
  Songaila, Tumlinson, Cowie, Deharveng, Friedman, Giroux, Green, Hutchings,
  Jenkins, Kruk, Moos, Morton, Sembach \& Tripp}{Kriss
  et~al.}{2001}]{Kriss2001}
Kriss G.~A.,  Shull J.~M.,  Oegerle W.,  Zheng W.,  Davidsen A.~F.,  Songaila
  A.,  Tumlinson J.,  Cowie L.~L.,  Deharveng J.~M.,  Friedman S.~D.,  Giroux
  M.~L.,  Green R.~F.,  Hutchings J.~B.,  Jenkins E.~B.,  Kruk J.~W.,  Moos
  H.~W.,  Morton D.~C.,  Sembach K.~R.,    Tripp T.~M.,  2001, \Science, 293,
  1112

\bibitem[\protect\citeauthoryear{{Leitherer}, {Schaerer}, {Goldader},
  {Delgado}, {Robert}, {Kune}, {de Mello}, {Devost} \& {Heckman}}{{Leitherer}
  et~al.}{1999}]{Leitherer}
{Leitherer} C.,  {Schaerer} D.,  {Goldader} J.~D.,  {Delgado} R.~M.,  {Robert}
  C.,  {Kune} D.~F.,  {de Mello} D.~F.,  {Devost} D.,    {Heckman} T.~M.,
  1999, \apjs, 123, 3

\bibitem[\protect\citeauthoryear{{Madau}, {Haardt} \& {Rees}}{{Madau}
  et~al.}{1999}]{Madau1999}
{Madau} P.,  {Haardt} F.,    {Rees} M.~J.,  1999, \apj, 514, 648

\bibitem[\protect\citeauthoryear{{Meiksin}}{{Meiksin}}{2005}]{Meiksin2005}
{Meiksin} A.,  2005, \mnras, 356, 596

\bibitem[\protect\citeauthoryear{{Miralda-Escud{\'e}}, {Haehnelt} \&
  {Rees}}{{Miralda-Escud{\'e}} et~al.}{2000}]{Miralda2000}
{Miralda-Escud{\'e}} J.,  {Haehnelt} M.,    {Rees} M.~J.,  2000, \apj, 530, 1

\bibitem[\protect\citeauthoryear{{Schirber} \& {Bullock}}{{Schirber} \&
  {Bullock}}{2003}]{Schirber2003}
{Schirber} M.,  {Bullock} J.~S.,  2003, \apj, 584, 110

\bibitem[\protect\citeauthoryear{{Sheth} \& {Tormen}}{{Sheth} \&
  {Tormen}}{1999}]{Sheth}
{Sheth} R.~K.,  {Tormen} G.,  1999, \mnras, 308, 119

\bibitem[\protect\citeauthoryear{{Smette}, {Heap}, {Williger}, {Tripp},
  {Jenkins} \& {Songaila}}{{Smette} et~al.}{2002}]{Smette2002}
{Smette} A.,  {Heap} S.~R.,  {Williger} G.~M.,  {Tripp} T.~M.,  {Jenkins}
  E.~B.,    {Songaila} A.,  2002, \apj, 564, 542

\bibitem[\protect\citeauthoryear{{Spergel}, {Bean}, {Dore}, {Nolta}, {Bennett},
  {Hinshaw}, {Jarosik}, {Komatsu}, {Page}, {Peiris}, {Verde}, {Barnes},
  {Halpern}, {Hill}, {Kogut}, {Limon}, {Meyer}, {Odegard}, {Tucker} \&
  {Weiland}}{{Spergel} et~al.}{2006}]{Spergel2006}
{Spergel} D.~N.,  {Bean} R.,  {Dore} O.,  {Nolta} M.~R.,  {Bennett} C.~L.,
  {Hinshaw} G.,  {Jarosik} N.,  {Komatsu} E.,  {Page} L.,  {Peiris} H.~V.,
  {Verde} L.,  {Barnes} C.,  {Halpern} M.,  {Hill} R.~S.,  {Kogut} A.,  {Limon}
  M.,  {Meyer} S.~S.,  {Odegard} N.,  {Tucker} G.~S.,    {Weiland} J.~L.,
  2006, astro-ph/0603449

\bibitem[\protect\citeauthoryear{{Stiavelli}, {Fall} \& {Panagia}}{{Stiavelli}
  et~al.}{2004}]{Stiavelli2004}
{Stiavelli} M.,  {Fall} S.~M.,    {Panagia} N.,  2004, \apjl, 610, L1

\bibitem[\protect\citeauthoryear{{Telfer}, {Zheng}, {Kriss} \&
  {Davidsen}}{{Telfer} et~al.}{2002}]{Telfer}
{Telfer} R.~C.,  {Zheng} W.,  {Kriss} G.~A.,    {Davidsen} A.~F.,  2002, \apj,
  565, 773

\bibitem[\protect\citeauthoryear{{Theuns}, {Bernardi}, {Frieman}, {Hewett},
  {Schaye}, {Sheth} \& {Subbarao}}{{Theuns} et~al.}{2002}]{Theuns}
{Theuns} T.,  {Bernardi} M.,  {Frieman} J.,  {Hewett} P.,  {Schaye} J.,
  {Sheth} R.~K.,    {Subbarao} M.,  2002, \apjl, 574, L111

\bibitem[\protect\citeauthoryear{{Thoul} \& {Weinberg}}{{Thoul} \&
  {Weinberg}}{1996}]{thoul}
{Thoul} A.~A.,  {Weinberg} D.~H.,  1996, \apj, 465, 608

\bibitem[\protect\citeauthoryear{{White}, {Becker}, {Fan} \& {Strauss}}{{White}
  et~al.}{2003}]{White2003}
{White} R.~L.,  {Becker} R.~H.,  {Fan} X.,    {Strauss} M.~A.,  2003, \aj, 126,
  1

\bibitem[\protect\citeauthoryear{{Wyithe} \& {Loeb}}{{Wyithe} \&
  {Loeb}}{2003a}]{Wyithe2003a}
{Wyithe} J.~S.~B.,  {Loeb} A.,  2003a, \apj, 586, 693

\bibitem[\protect\citeauthoryear{{Wyithe} \& {Loeb}}{{Wyithe} \&
  {Loeb}}{2003b}]{WyitheSMBH2003}
{Wyithe} J.~S.~B.,  {Loeb} A.,  2003b, \apj, 595, 614

\bibitem[\protect\citeauthoryear{{Wyithe} \& {Loeb}}{{Wyithe} \&
  {Loeb}}{2006}]{Wyithe2006}
{Wyithe} J.~S.~B.,  {Loeb} A.,  2006, \nat, 441, 322

\bibitem[\protect\citeauthoryear{{Yan} \& {Windhorst}}{{Yan} \&
  {Windhorst}}{2004}]{Yan2004}
{Yan} H.,  {Windhorst} R.~A.,  2004, \apjl, 600, L1

\end{thebibliography}
\end{document}